# Nanorods based on mesoporous silica containing iron oxide nanoparticles as catalytic nanomotors: study of motion dynamics

Rafael Mestre[a], Núria Cadefau[a], Ana C. Hortelão[a], Jan Grzelak[b], Martí Gich[b], Anna Roig*[b], Samuel Sánchez*[a,c]

[a]   R. Mestre, N. Cadefau, A. C. Hortelão, Prof. S. Sánchez
      Institute for Bioengineering of Catalonia (IBEC), The Barcelona Institute of Science and Technology (BIST).
      Baldiri-Reixac 10-12, 08028 Barcelona, Spain.
      E-mail: ssanchez@ibecbarcelona.eu

[b]   J. Grzelak, Dr. M. Gich, Prof. A. Roig.
      Institut de Ciència de Materials de Barcelona (ICMAB-CSIC).
      Campus UAB, 08193 Bellaterra, Catalonia, Spain.
      E-mail: roig@icmab.es

[c]   Prof. S. Sánchez
      Institució Catalana de Recerca i Estudis Avançats (ICREA).
      Passeig de Lluís Companys 23, 08010 Barcelona, Spain.
      E-mail: ssanchez@ibecbarcelona.eu

**Abstract:**

Self-propelled particles and, in particular, those based on mesoporous silica, have raised considerable interest due to their potential applications in the environmental and biomedical fields thanks to their biocompatibility, tunable surface chemistry and large porosity. Although spherical particles have been widely used to fabricate nano- and micromotors, not much attention has been paid to other geometries, such as nanorods. Here, we report the fabrication of self-propelled mesoporous silica nanorods (MSNRs) that move by the catalytic decomposition of hydrogen peroxide by a sputtered Pt layer, $Fe_2O_3$ nanoparticles grown within the mesopores, or the synergistic combination of both. We show that motion can occur in two distinct sub-populations characterized by two different motion dynamics, namely enhanced diffusion or directional propulsion, especially when both catalysts are used. These results open up the possibility of using MSNRs as chassis for the fabrication of self-propelled particles for the environmental or biomedical fields.

## Introduction

Self-propelled nano- and microparticles, inspired by the motion of biological organisms, have emerged as a new and interesting type of active systems which have found numerous applications in fundamental physics[1–4], biomedicine[5–7] or environmental applications[8]. These systems can perform active motion by the conversion of chemical energy from the environment into kinetic energy and are commonly described under the theory of active Brownian motion[9,10].

Multiple types of self-propelled particles, also called micro- and nanomotors or swimmers, have been reported over the last fifteen years[1,2,5]. In particular, chemically powered swimmers harness free energy from the surrounding medium and convert it into motion.[11,12] Initial examples of catalytic micromotors were based on bimetallic rod-like structures, which were able to self-propel due to the decomposition of hydrogen peroxide ($H_2O_2$) *via* electrokinetic mechanisms[13–17]. Likewise, other motors based on carbon fibers[18], nanotubes[19,20] or microtubes[21,22] were propelled by enzyme bio-catalysis. The motion of this type of asymmetric swimmers based on nano- or microtubular structures mainly relies on the production of $O_2$ or $H_2$, which nucleate into bubbles within the tubular cavities and provoke displacement by exiting in a jet-like fashion[19,21,23–25]. Several proof-of-concept applications have been demonstrated with these microjets such as the transport of cargo on-chip[26–29], cleaning of contaminated water[8,22,30–32] or biosensing[33–38].

All the tubular structures described so far have reported a heterogeneous library of scaffolds with a wide range of applications, but there is a need of more versatile materials with tunable properties and the possibility of easily loading cargo or drugs. On this subject, silica has become an excellent candidate for nano- and micromotor applications[39,40] due to their ease of fabrication in different sizes and architectures[5,41–43], their surface chemistry that allows the attachment of biomolecules like enzymes or antibodies[44,45], as well as the possibility of obtaining pores of different sizes to load drugs or other molecules within them[45,46].

Here, we explore the use of mesoporous silica nanorods (MSNRs) as chassis for the development of novel non-spherical self-propelled silica-based systems. To the best of our knowledge, all of the nano- and micromotors based on mesoporous silica use spherical nano- or microparticles or hollow nano- or microjets, and none have explored the possibility of using nanorods with cylindrical pores oriented along the rod axis, which could be used to load drugs or other molecules. In our case, as a proof of concept, we incorporated iron oxide nanoparticles inside the pores (MSNR-Fe), to aid the motion via Fenton-like reactions with $H_2O_2$ fuel. Furthermore, we partially coated the MSNRs, with and without $Fe_2O_3$ nanoparticles, with Pt to create Janus-like structures (MSNR-Fe-Pt and MSNR-Pt), introducing an additional level of asymmetry. Results showed that the reaction catalyzed by the internal $Fe_2O_3$ nanoparticles alone (MSNR-Fe) is not enough to yield efficient active motion, but its coupling with Pt deposition enhances its effect (MSNR-Fe-Pt), even beyond those with only partial Pt coverage without nanoparticles (MSNR-Pt). Moreover, our analyses revealed two types of sub-populations with distinct types of motion, one of them characterized by an enhanced Brownian motion, while the other shows enhanced directionality and propulsive behavior, pointing towards the possibility of modulating the motion behavior by fine-tuning the fabrication strategy.

## Experimental Section

*Synthesis of MSNRs*

MSNRs were synthesized by a sol-gel method that used silica alkoxide as a precursor and Pluronic P123 as a surfactant by a modified procedure from ref. [47]. Briefly, 16.5 g of hydrochloric acid (37% wt/v) was diluted in 81 mL of Milli-Q water and 2 g of Pluronic P123 were subsequently dissolved in it. The mixture was heated up at 40 °C with vigorous stirring of 700 rpm for 3 h. Then, 4.24 g of tetraethyl orthosilicate (TEOS) were added to the previous mixture and stirred for 5 min. The reaction was left to continue for 24 h at 40 °C. Then, the mixture was transferred to an oven at 80 °C for another 24 h. Afterward, the synthesized silica nanorods were filtered and dried at 50 °C overnight. Finally, the MSNRs were washed in ethanol in a Soxhlet extractor (24 cycles) to remove the surfactant and they were annealed at 550 °C for 5 h.

*Synthesis of MSNR-Fe*

As in ref. [47], iron oxide ($Fe_2O_3$) nanoparticles were synthesized inside the pores of MSNRs using a wet impregnation method. MSNRs and iron (III) nitrate nonahydrate were mixed and heated above the melting temperature of the iron (III) nitrate nonahydrate precursor (47.2 °C). Then, thermal decomposition of the precursor was carried out at 425 °C for 2 h.

*Fabrication of MSNR-Pt and MSNR-Fe-Pt*

To fabricate nanomotors based on Pt, monolayers of MSNR and MSNR-Fe were deposited by adding 50 µL of a solution (2 mg/mL) of each type of nanorod on a glass slide of 24x24 mm kept at 45°, which had been previously cleaned with acetone and isopropanol and treated with oxygen plasma. After the MSNR monolayers had dried, Pt was sputtered on the exposed side of the MSNR using a Leica EM ACE600 sputter, at a rate of 0.4 nm/s, with a working distance of 45 mm, a current of 35 mA and achieving a thickness of 10 nm of sputtered Pt.

*Characterization of the three types of MSNRs*

The MSNRs and the three types of motors derived from them were characterized by SEM, TEM, DLS and EDX mapping. SEM imaging was performed using a FEI NOVA NanoSEM 230 scanning microscope at 10 kV. TEM imaging was performed with a JEOL JEM-2100 transmission microscope. The electrophoretic mobility of the nanorods was obtained with a Wyatt Möbius DLS machine coupled with an Atlas pressurization system, using a 532 nm wavelength laser and a detector angle of 163.5°. EDX mapping was performed with a Magellan 400L Field Emission Scanning Electron Microscope with attached X-Max Ultim Extreme EDX (Oxford Instruments). The crystalline phase of iron oxide particles was analyzed in a JEOL JEM-1210 TEM operating in diffraction mode, at an operating voltage of 120 kV. One drop of MSNR dispersion in ethanol was placed on a TEM grid (Micro to Nano, EMR carbon support film on copper 200 square mesh).

*Optical video recording of MSNRs*

The motion of the nanorods and nanomotors was performed with an optical inverted microscope (Leica DMi8), equipped with a 63x water immersion objective (NA = 1.2). Briefly, a concentrated solution of nanomotors was diluted in 1 mL of water. Then, a drop of this solution was mixed in a 1:1 ratio with a $H_2O_2$ solution to achieve final concentrations of 0%, 0.5%, 1%, 1.5%, 3% and 5% (v/v), and the final solution was placed in a glass coverslip under the microscope. Several videos of 30 s at 25 frames per second (FPS) were recorded for each condition. The trajectories of the particles were obtained with a previously available home-made tracking software based on Python that locates and tracks the particles based on computer vision techniques, returning the positions of the particles with time[20,48]. For each of the conditions of nanomotors and fuel concentration, a sample size between N = 50 - 150 was used. The 2D-projected mean squared displacement (MSD) was computed from the particle's trajectories in time $(x(t), y(t))$ as:

$$MSD(\Delta t) = <\left(x(t_0 + \Delta t) - x(t_0)\right)^2> +$$
$$+ <\left(y(t_0 + \Delta t) - y(t_0)\right)^2>,$$

where $<\cdot>$ represents ensemble and time average, $t_0$ is the initial time and $\Delta t$ is the elapsed time or the inverse of the FPS. The first 2 s of the MSD were fitted to an equation of the form:

$$MSD(\Delta t) = D\Delta t^\alpha,$$

from which the α exponent was extracted. Based on the value of α, different approximations were performed. For $\alpha > 1.2$, the MSD was assumed to describe a propulsive behavior which depends on time as a combination of $\Delta t$ and $\Delta t^2$ polynomials, as in the well-known formula[10]:

$$MSD(\Delta t) = 4D_t\Delta t + v^2\Delta t^2,$$

where $D_t$ is the translational diffusion coefficient and $v$ is the propulsive speed. When $\alpha < 1.2$, the nanomotors were assumed to behave with enhanced Brownian motion and were fitted to the equation:

$$MSD(\Delta t) = 4D_e\Delta t,$$

where $D_e$ is the enhanced diffusion coefficient, higher than $D_t$, as it contains the information about the reaction activity.

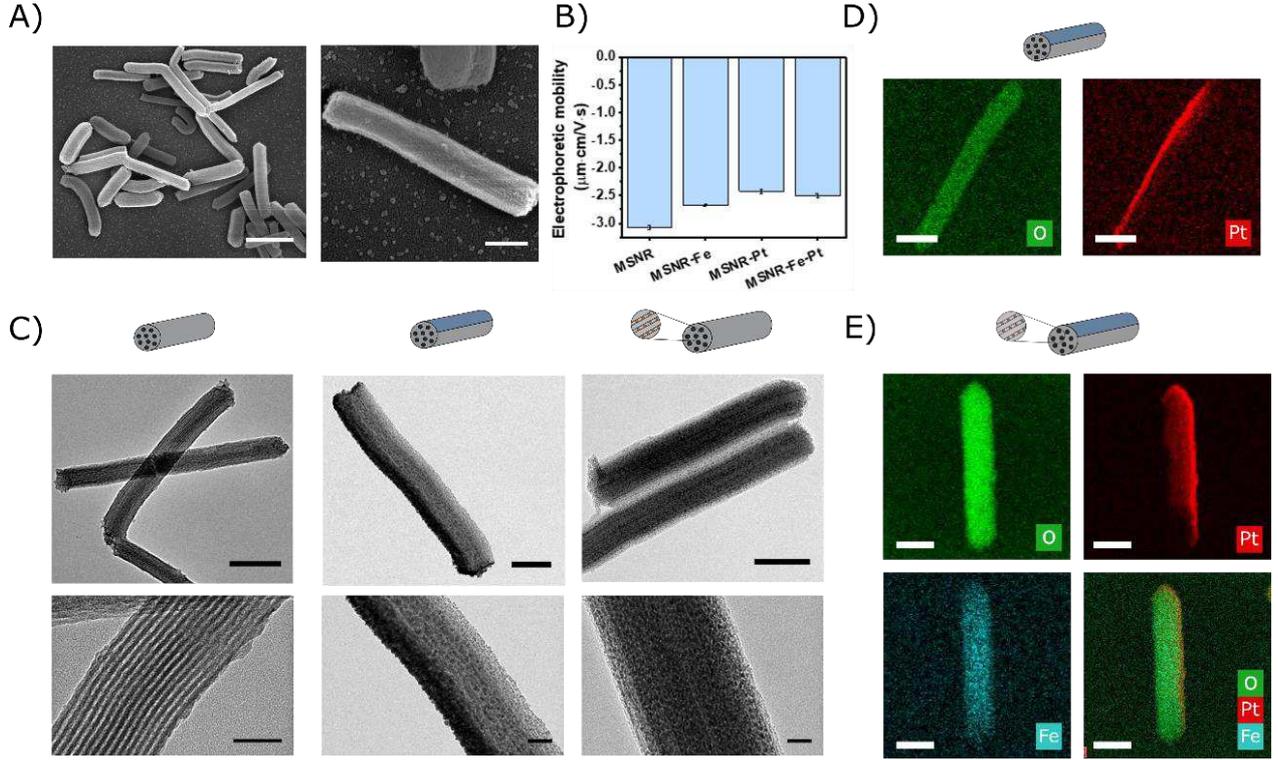

**Figure 1.** Characterization of the MSNRs motors under SEM, TEM, DLS and EDX. A) SEM images of MSNRs. Scale bars: 1 µm (left) and 250 nm (right). B) Electrophoretic mobility of different types of MSNRs motors. C) TEM imaging of MSNRs (left), MSNRs sputtered with Pt (MSNRs-Pt) at center and MSNRs with iron oxide nanoparticles inside its mesopores (MSNRs-Fe) at the right. Scale bars: 250 nm (top images) and 50 nm (bottom images). D) EDX mapping of MSNRs-Pt and E) MSNRs sputtered with Pt and iron oxide nanoparticles in its mesopores (MSNRs-Fe-Pt). Scale bars: 250 nm.

## Results and Discussion

Rod-shaped silica nanomotors, self-propelled by inorganic catalytic reactions, were synthesized by a sol-gel method that uses Pluronic P123 as surfactant and yields nanorods with straight cylindrical hexagonally ordered mesopores along the rods main axis (MSNRs). These nanorods were coupled with one or more metallic components that can catalyze reactions and become nanomotors in three different manners: i) nanorods horizontally placed on a substrate sputtered with Pt to achieve one-side asymmetric metallic coating (MSNR-Pt); ii) nanorods with $Fe_2O_3$ nanoparticles trapped inside the mesopores (MSNR-Fe); and iii) a combination of the two approaches, namely nanorods with Pt on the surface and $Fe_2O_3$ nanoparticles inside the pores (MSNRs-Fe-Pt). Pt is one of the most used metals to catalyze the decomposition of $H_2O_2$ to generate motion of nano- and micro-structures and it generally needs an asymmetric covering for an efficient motion[49]. Sputtering Pt on MSNRs produces an asymmetric coverage of half of the rod surface, in a longitudinal manner, which might seem insufficient to induce directionality. One could expect that such a lateral coverage would increase the diffusivity of the nanorods by increasing their active fluctuation, covering a wider area in an enhanced Brownian motion. To improve the directionality of the motion, $Fe_2O_3$ nanoparticles were synthesized within the mesopores, similarly to a previously reported strategy where enzymes were added inside micro-rods to achieve directional motion[20]. Thanks to the reduction of $Fe^{3+}$ to $Fe^{2+}$ and the oxidation of $Fe^{2+}$ to $Fe^{3+}$ in a Fenton reaction, active motion could be induced due to the expulsion of the reaction products through the pores. Finally, from the combination of both species (Pt and $Fe_2O_3$) one could expect an even more efficient and directional motion.

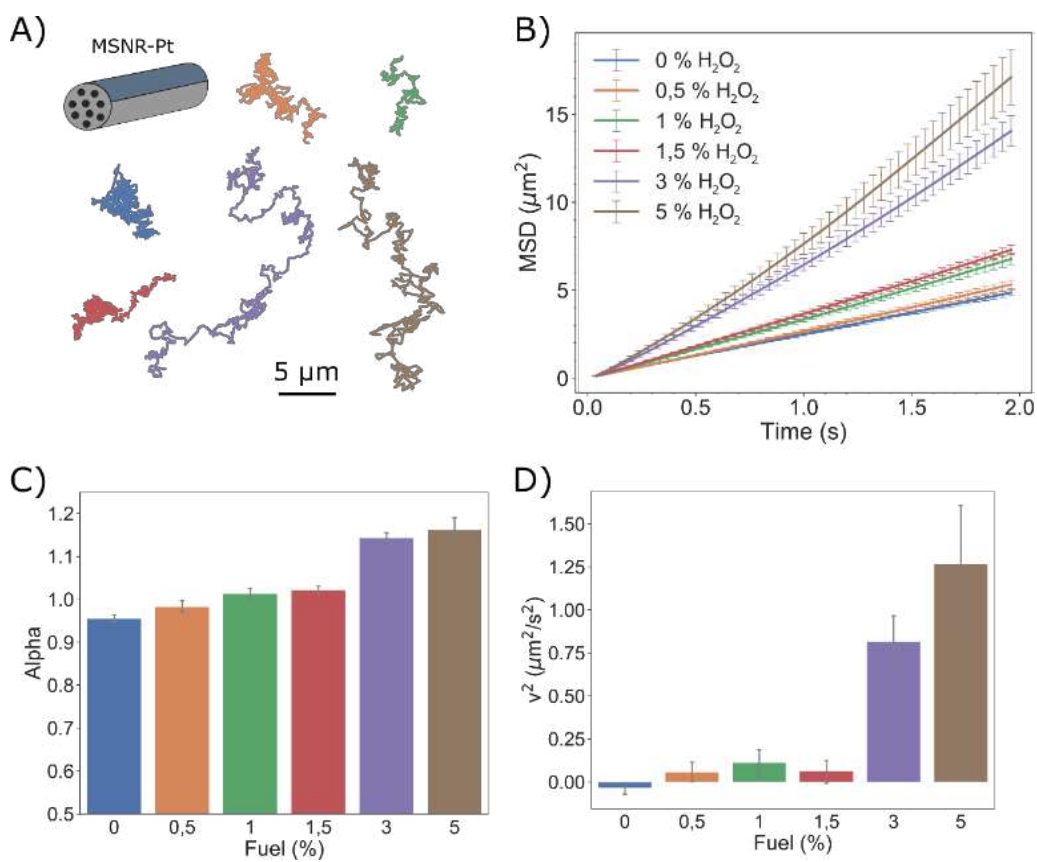

**Figure 2.** Analysis of motion of MSNR-Pt. A) Representative 30 s trajectories of the motors under different fuel ($H_2O_2$) concentrations and B) their corresponding MSDs. C) Performing a logarithmic fitting reveals that the exponent α increases with fuel concentration indicating the presence of superdiffusive regime. D) Performing a quadratic fitting instead allows extracting the squared speed of the motors, showing only a modest positive value for high fuel concentrations.

The three types of motors were characterized by scanning electron microscopy (SEM), transmission electron microscopy (TEM), dynamic light scattering (DLS) and energy-dispersive X-ray (EDX) spectroscopy mapping (**Figure 1**). Figure 1A shows SEM images of the MSNRs. An average length of 1.4 ± 0.3 µm and a diameter of 0.30 ± 0.05 µm has been obtained by analyzing 300 nanorods from TEM images. The electrophoretic mobility of Figure 1B shows that, as expected from the hydroxyl-terminated silica, the MSNRs and the three types of nanomotors considered, have a negatively charged surface, demonstrating that the addition of the metallic components did not greatly alter the surface charge. Figure 1C shows TEM images of MSNRs (left), MSNRs-Pt (center) and MSNRs-Fe (right). The longitudinal mesopores of MSRNs, with diameters of about 5-6 nm, are clearly visible at high magnification. For MSNR-Pt, the partial coverage of Pt can be also observed, displaying a darker contrast in the Pt-sputtered region. Finally, in MSNR-Fe the mesopores are less visible, but the speckle-like pattern of darker spots is attributed to the presence of $Fe_2O_3$ nanoparticles, as revealed by the electron diffraction pattern presented in Figure S1. The diffraction rings were used for indexation of maghemite crystallographic planes (P4$_1$32 space group)[47]. Higher intensity spots along the diffraction rings show the preferred orientation of maghemite particles. EDX mapping further demonstrates the previous statements. Figure 1D shows the distribution of oxygen (indicating the presence of silica) and Pt, clearly revealing its presence on half of the surface. Figure 1E, which presents the EDX mapping of MSRNs-Fe-Pt, also displays this, plus the presence of Fe all along the nanorod, in agreement with the presence of Fe$_2$O$_3$ nanoparticles inside the MSNRs.

The motion of MSNR-Pt under the presence of hydrogen peroxide (H$_2$O$_2$) at different volume concentrations is shown in **Figure 2.** Both the representative trajectories and the MSDs clearly show that the area covered by the nanomotors increases with fuel concentration (Figures 2A and 2B). For a nanoparticle of a non-spherical shape, such as a nanorod, the translational and rotational diffusions are coupled in the laboratory's frame of reference, as the viscous drag forces depend on the orientation of the particle. In a frame of reference located on the nanorod, these diffusivities can be decoupled from rotations, displaying greater translational diffusion along the rod main axis. However, as optical microscopy does not allow to precisely identify the orientation of heavily fluctuating nanorods, we cannot study the diffusion along different directions separately, and we need to take a global approach, considering the MSNRs as point particles which follow the Einstein-Smoluchowski equation of diffusion. Nonetheless, when fuel is added to a cylindrical nanomotor, it is advisable to study its motion dynamics using a more general formula, assuming $MSD(\Delta t) = D\Delta t^\alpha$. Figure 2C shows the α values after a fitting to the previous equation. All fuel concentrations up to 1.5% show roughly $\alpha = 1$, indicating normal diffusive behavior. However, it can be seen that, while the MSD remained linear, the diffusions of the MSRN-Pt were enhanced due to the catalytic activity since for 3% and 5% of fuel concentration $\alpha$ takes values higher than 1, indicating superdiffusive behavior. In general, nano- and micro-motors showing superdiffusion follow the well-known equation of the form $MSD(\Delta t) = 4D_t\Delta t + v^2\Delta t^2$, where $v$ is the propulsive speed of the particle and $D_t$ its translational diffusion. Depending on the ratio between these two terms, a fitting of the type $\sim \Delta t^\alpha$ to the MSD will not yield $\alpha = 1$ or $\alpha = 2$, but a value in between, as both the linear and quadratic terms contribute to the curvature. Therefore, if we assume that these MSNR-Pt can achieve a propulsive regime with $\alpha = 2$, we can perform a fitting to the quadratic equation and extract the propulsive speed, as reported in Figure 2D. In this case, and throughout this work, the squared value of the propulsive speed is reported, as cases with $v = 0$ can yield negative squared speeds, which are physically meaningless. It can be seen that higher fuel concentrations show positive speed, although their values barely reach 1 µm/s$^2$ and their variability is large. This difference seems small compared to the very clear differences between fuels observed in the representative trajectories of Figure 2A and their corresponding MSDs, suggesting that a more in-depth analysis should be considered.

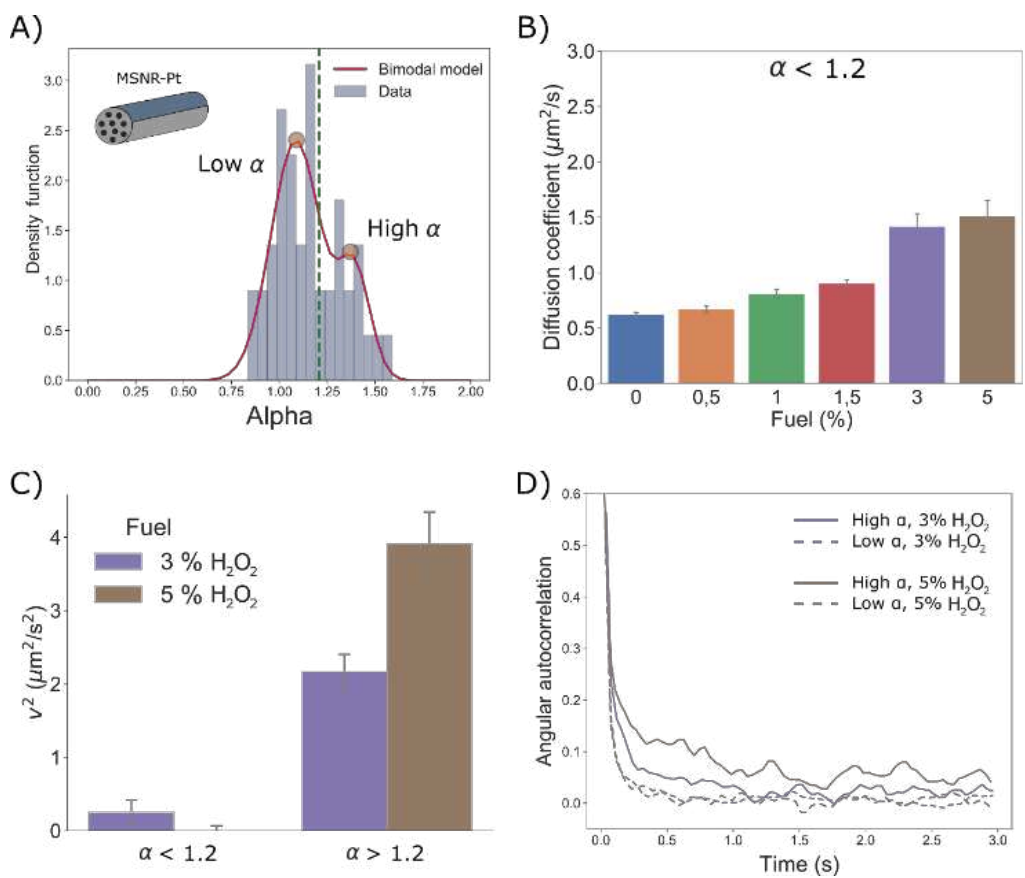

**Figure 3.** Analysis of subpopulations of MSNRs-Pt. A) The distribution of MSD exponents (α) for a fuel concentration of 5% reveals that there are two subpopulations of motors, some behaving diffusively ($\alpha = 1$) and others superdiffusively ($\alpha > 1$). B) Approximating all motors with $\alpha < 1.2$ to a diffusive behavior with $\alpha = 1$ shows that for higher fuel concentrations, there is an enhanced diffusion of the motors. C) Approximating the MSDs to a quadratic equation reveals that the sub-population of motors behaving superdiffusively have a squared speed roughly four times higher than the one reported in the global analysis; those in the low α sub-population have zero propulsive speed. D) The angular autocorrelation functions of both sub-populations for higher fuel concentrations show that the high α sub-population has a higher directionality than the low α one, which shows no correlation.

A closer study to the distribution of α values over the whole population of nanomotors can improve the analysis. **Figure 3A** shows the probability density function of all α for a fuel concentration of 5%. The histogram corresponding to the density function shows two distinct peaks, which indicates the presence of two sub-populations with different average α, resulting in a bimodal distribution. The first peak represents nanomotors that move diffusively with α ~ 1, but might have an enhanced diffusion thanks to the catalytic activity. The second peak of higher α represents a sub-population of nanomotors that move in a superdiffusive manner, and probably with more directionality. It is clear, then, that we cannot treat these two populations with the same statistical approach, as they represent different motion types. The presence of a bimodal distribution is more clearly seen at high fuel concentrations, although for moderate concentrations a spread in the standard deviation of this distribution, indicating an increase in the overall activity of the nanomotors, can also be observed (Figure S2). This effect might be due to the imperfect control of the Pt coating, giving rise to two different types of coverage asymmetries each one favoring a specific kind of motion, especially for 5% $H_2O_2$. For this reason, we decided to set a threshold of α = 1.2 for the statistical treatment of the nanomotors. On the one hand, if α < 1.2, a nanomotor was considered as moving diffusively and approximated to α = 1 to extract the translational diffusivity. On the other hand, if α > 1.2, the nanomotors were assumed to move propulsively and fitted to a quadratic equation to extract their propulsive speed.

Figure 3B shows the extracted diffusion coefficient of the sub-population of MSNR-Pt with α < 1.2. As expected, even if their motion is diffusive, the diffusivities were enhanced compared to the control case (0% $H_2O_2$), reaching up to a 3-fold increase due to the catalytic reaction of Pt with $H_2O_2$. Performing a quadratic fitting to the MSDs of the two highest concentrations (which clearly show this sub-population division), we can see that the sub-populations of motors with α < 1.2 do not have propulsive speed, as expected, but those with α > 1.2 do show significant speed. Compared with the results of Figure 2D, we can see that the low speed reported in that case was a result of the mixing up of the two sub-populations, further justifying the approach of separating both sub-populations, as they show different dynamics. Indeed, the probability density of squared speeds of the whole population of nanomotors (Figure S5) also reveals two distinct sub-populations, one with an average of 0 $\mu m^2/s^2$, corresponding to nanomotors presenting enhanced diffusion. Finally, the angular autocorrelation presented in Figure 3D demonstrates that the high-α sub-population has a more directional motion. Dashed lines, corresponding to low-α cases, show a lack of autocorrelation typical of Brownian fluctuations, while the continuous lines, corresponding to high-α nanomotors, display a long-term autocorrelation, especially for the 5% case.

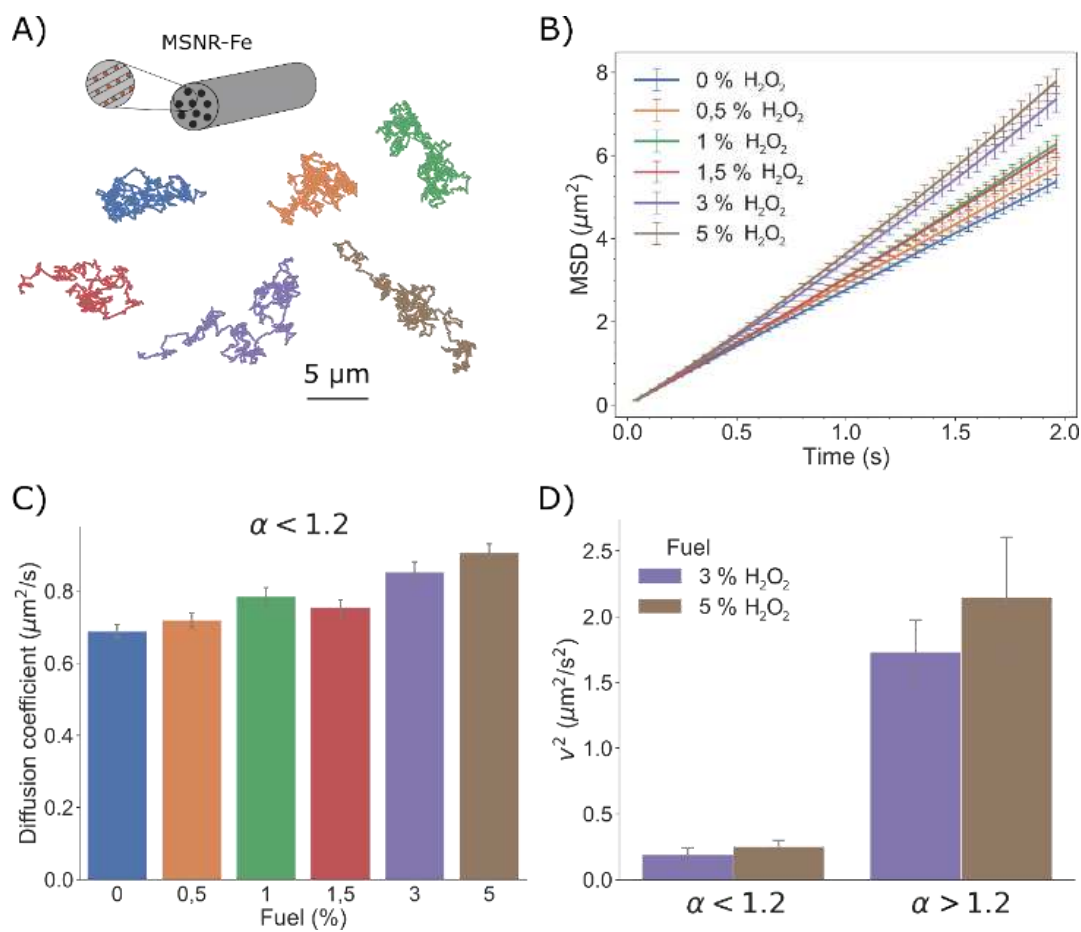

**Figure 4.** Analysis of motion of MSNRs-Fe. A) Representative 30 s trajectories of the motors under different fuel concentrations and B) their corresponding MSDs. C) Approximating all motors with α < 1.2 to a diffusive behavior with α = 1 shows a very modest increase of the diffusion coefficient with fuel concentration. D) Approximating both sub-populations to a quadratic equation shows that only those with α > 1 have a small positive speed squared.

**Figure 4** shows the same type of analysis for the MSNR-Fe nanomotors. The representative trajectories and the MSDs of Figures 4A and 4B already indicate modest changes when increasing the concentration of fuel. Indeed, a linear fitting to the MSD shows small increases of the diffusion coefficient compared to the control case. The probability density function of α does not show such a distinct separation of sub-populations with different dynamics, although for 5% $H_2O_2$ there is a slight deviation from a null average speed (Figure S3). Indeed, a linear fitting (Figure 4C) shows a small increment of the diffusion coefficient for higher fuel concentrations, although a quadratic fitting yields only a moderate increment of the speed for 3% and 5% $H_2O_2$ (Figure 4D and S6). Therefore, the presence of $Fe_2O_3$ nanoparticles inside the mesopores seems to slightly increase the diffusivity of the nanorods, but not in a strong and directional manner as initially hypothesized.

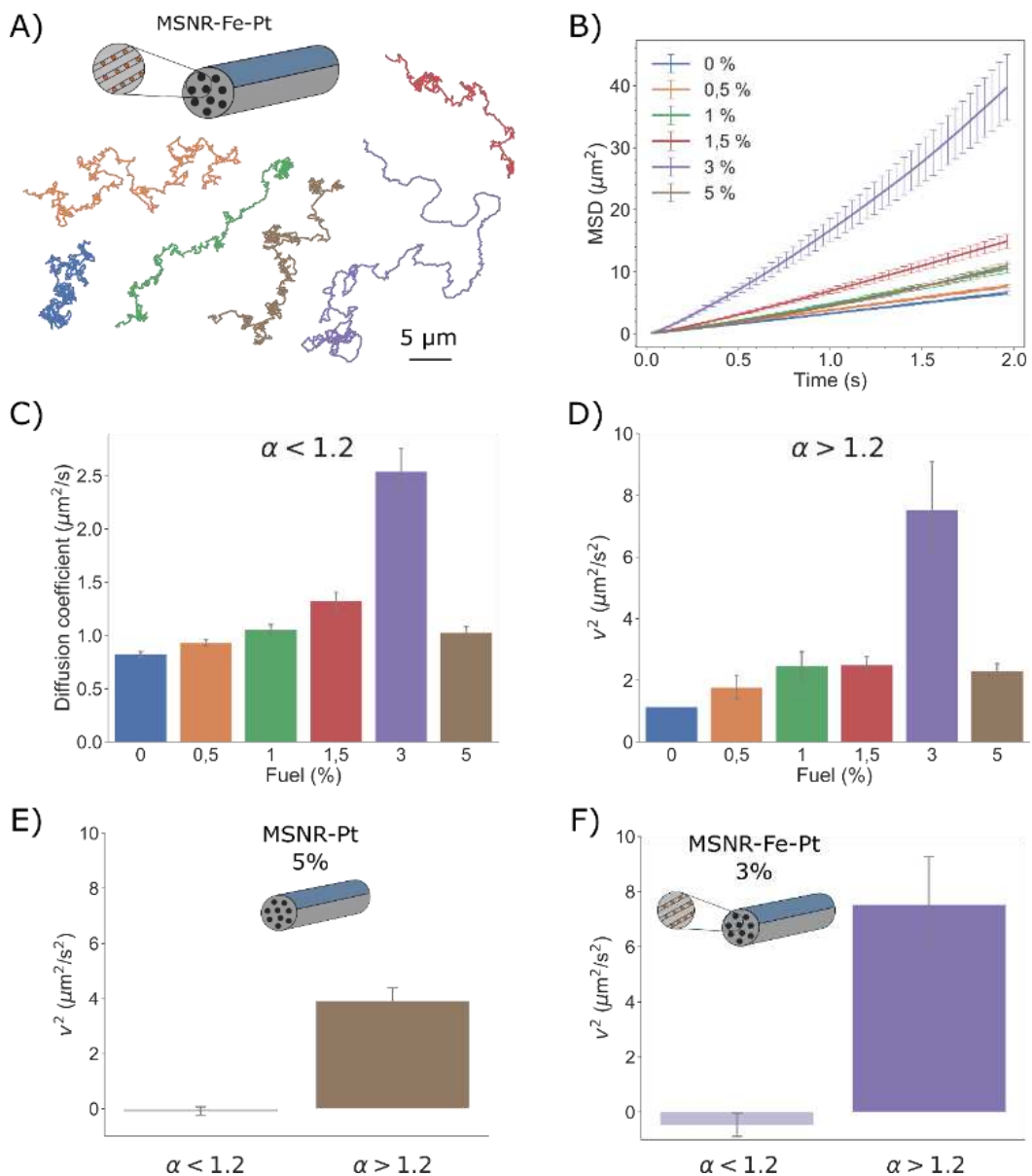

**Figure 5**. Analysis of motion of MSNRs-Fe-Pt. A) Representative 30 s trajectories of the motors under different fuel concentrations and B) their corresponding MSDs. C) Approximating the sub-population with α < 1.2 to a diffusive behavior with α = 1 shows the highest increase of enhanced diffusion coefficient for a concentration of 3%. D) Approximating the sub-population with α > 1.2 to a quadratic equation shows the same behavior, with positive squared speeds even for low fuel concentrations (notice that the positive value for 0% corresponds to a single outlier that could be most likely due to statistical fluctuations because of MSD variability). E-F) Comparing the speed squared of the best concentration of the MSNR-Pt motors (5%) with the one of MSNR-Fe-Pt (3 %) shows that the combination of both metallic materials yields a two-fold increase in speed.

Finally, **Figure 5** presents the analysis of MSNR-Fe-Pt nanomotors, using a combination of both catalytic mechanisms. Figures 5A and 5B show directional trajectories for all of the fuel concentrations and clear differences in the MSDs, which can cover in 2 s a much wider range than for the previous systems. Indeed, the diffusion coefficient of the low-α sub-population and the squared speed of the high-α sub-population display the same trend, with increasing values for higher fuel concentrations, presenting a significant peak for a fuel concentration of 3% (Figure

S4 and S8). It is striking that such an increment occurs for this concentration and not for the highest one, but some inhibition mechanism related to increased reactivity (as both catalysts, Pt and $Fe_2O_3$, are present) could be taking place since, for example, the diffusion for 5% is lower than for 1.5%. Regardless, motion for 3% of fuel concentration is highly efficient, yielding high speeds for directional motors and a greatly enhanced diffusion for non-directional ones. Figure 5E and 5F compare the best two cases of propulsive motion for MSNR-Pt (with 5% of fuel) and MSNR-Fe-Pt (with 3% of fuel). For low-α sub-populations, there is no speed, as expected, while for the high-α sub-populations, MSNR-Fe-Pt show a two-fold increase in squared speed compared to MSNR-Pt. We hypothesize that the presence of the Pt half-coverage can create some directional motors (as discussed in Figure 2 and reported in 5E), a behavior that is enhanced by the presence of $Fe_2O_3$ nanoparticles inside the pores. $Fe_2O_3$ nanoparticles, by themselves, do not produce directional motion (as discussed in Figure 4), but the directional symmetry breaking caused by the motion greatly enhances the speed of the MSRN-Pt-Fe. This synergistic effect could be explained by a stabilization effect after the expulsion of the Fenton reaction products through the pores of the MSNRs. Due to the inherent symmetry of the rod structure, only when motion is already induced by the presence of the partial Pt coverage, inversion of the trajectory is inhibited by this stabilization effect, achieving more directional motion.

## Conclusion

Mesoporous silica is an excellent chassis for micro- and nanomotors due to its convenient surface chemistry, biocompatibility and possibility of encapsulating drugs or other molecules. In this work, we report the fabrication and characterization of non-spherical nanomotors based on MSNRs with two different catalytic engines, namely partial surface coverage with Pt of the nanorods and $Fe_2O_3$ nanoparticles inside its pores. Upon addition of $H_2O_2$ as fuel, two sub-populations of nanomotors with different motion dynamics arose, likely due to a lack of fine geometrical control during the Pt deposition. By studying separately these two sub-populations with different scaling exponent, "high α" and "low α", we conveyed that the low-α population is better described as moving by enhanced diffusion and the high α-population by propulsive and directional motion. Note that the motion of MSNR-Fe due to the Fenton reaction catalyzed by the $Fe_2O_3$ nanoparticles was not efficient to produce significant motion. However, when MSNR-Fe were coupled with partial coverage of Pt, both the enhanced diffusion and propulsive speed were greatly increased. This synergistic effect could be due to a stabilization of the nanorods from turning due to the expulsion of reaction products through the pores, enhancing the directionality of the motion[31,50–52]. All in all, we have demonstrated that MSNRs show high potential as chassis for the development of nanomotors with tunable motion dynamics in applications related to water remediation, due to the presence of Fenton reactions as well as in applications related to biomedicine, for which further research is needed to integrate more biocompatible power sources.

## Acknowledgements

R.M. thanks "la Caixa" Foundation through IBEC International PhD Programme "la Caixa" Severo Ochoa fellowships (code LCF/BQ/SO16/52270018). A.C.H. thanks MINECO for the Severo Ochoa PhD fellowship. J.G. has received financial support through the "la Caixa" INPhINIT Fellowship Grant for Doctoral Studies at Spanish Research Centers of Excellence (Grant code: LCF/BQ/DI17/11620041), "la Caixa" Banking Foundation (ID100010434), Barcelona, Spain. The CERCA program by the Generalitat de Catalunya, the Secretaria d'Universitats i Recerca del Departament d'Empresa i Coneixement de la Generalitat de Catalunya through the projects 2017SGR1148 & 2017SGR765 are acknowledged, as well as the Ministerio de Ciencia, Innovación y Universidades (MCIU) / Agencia Estatal de Investigación (AEI) / Fondo Europeo de Desarrollo Regional (FEDER, UE) through the projects RTI2018-098164-B-I00 & RTI2018-096273-B-I00 and Severo Ochoa Programme for Centres of Excellence in R&D (SEV-2015-0496). This project was also partially funded by Agencia Estatal de Investigación (CEX2018-000789-S). This project has received funding from the European Research Council (ERC) under the European Union's Horizon 2020 research and innovation programme (grant agreement No 866348).

## References

[1] S. Sanchez, L. Soler, J. Katuri, *Angew. Chemie - Int. Ed.* **2015**, *54*, 1414.

[2] J. Katuri, X. Ma, M. M. Stanton, S. Sánchez, *Acc. Chem. Res.* **2017**, *50*, 2.

[3] M. N. Popescu, W. E. Uspal, A. Domínguez, S. Dietrich, *Acc. Chem. Res.* **2018**, *51*, 2991.

[4] H. Stark, *Acc. Chem. Res.* **2018**, *51*, 2681.

[5] T. Patiño, X. Arqué, R. Mestre, L. Palacios, S. Sánchez, *Acc. Chem. Res.* **2018**, *51*, 2662.

[6] I. Ortiz-Rivera, M. Mathesh, D. A. Wilson, *Acc. Chem. Res.* **2018**, *51*, 1891.

[7] J. Wang, Z. Xiong, J. Zheng, X. Zhan, J. Tang, *Acc. Chem. Res.* **2018**, *51*, 1957.

[8] J. Parmar, D. Vilela, K. Villa, J. Wang, S. Sánchez, *J. Am. Chem. Soc.* **2018**, *140*, 9317.

[9] P. Romanczuk, M. Bär, W. Ebeling, B. Lindner, L. Schimansky-Geier, *Eur. Phys. J. Spec. Top.* **2012**, *202*, 1.

[10] J. R. Howse, R. A. L. Jones, A. J. Ryan, T. Gough, R. Vafabakhsh, R. Golestanian, *Phys. Rev. Lett.* **2007**, *99*, 048102.

[11] K. K. Dey, F. Wong, A. Altemose, A. Sen, *Curr. Opin. Colloid Interface Sci.* **2016**, *21*, 4.

[12] S. Sanchez, L. Soler, J. Katuri, *Angew. Chemie - Int. Ed.* **2015**, *54*, 1414.

[13] W. F. Paxton, K. C. Kistler, C. C. Olmeda, A. Sen, S. K. St. Angelo, Y. Cao, T. E. Mallouk, P. E. Lammert, V. H. Crespi, *J. Am. Chem. Soc.* **2004**, *126*, 13424.

[14] Y. Wang, R. M. Hernandez, D. J. Bartlett, J. M. Bingham, T. R. Kline, A. Sen, T. E. Mallouk, *Langmuir* **2006**, *22*, 10451.

[15] S. Fournier-Bidoz, A. C. Arsenault, I. Manners, G. A. Ozin, *Chem. Commun.* **2005**, 441.

[16] R. Laocharoensuk, J. Burdick, J. Wang, *ACS Nano* **2008**, *2*, 1069.

[17] U. K. Demirok, R. Laocharoensuk, K. M. Manesh, J. Wang, *Angew. Chemie* **2008**, *120*, 9489.

[18] N. Mano, A. Heller, *J. Am. Chem. Soc.* **2005**, *127*, 11574.

[19] D. Pantarotto, W. R. Browne, B. L. Feringa, *Chem. Commun.* **2008**, 1533.

[20] X. Ma, A. C. Hortelao, A. Miguel-López, S. Sánchez, *J. Am. Chem. Soc.* **2016**, *138*, 13782.

[21] S. Sanchez, A. A. Solovev, Y. Mei, O. G. Schmidt, *J. Am. Chem. Soc.* **2010**, *132*, 13144.

[22] J. Orozco, V. García-Gradilla, M. D'Agostino, W. Gao, A. Cortés, J. Wang, *ACS Nano* **2013**, *7*, 818.

[23] A. A. Solovev, Y. Mei, E. B. Ureña, G. Huang, O. G. Schmidt, *Small* **2009**, *5*, 1688.

[24] A. A. Solovev, W. Xi, D. H. Gracias, S. M. Harazim, C. Deneke, S. Sanchez, O. G. Schmidt, *ACS Nano* **2012**, *6*, 1751.

[25] V. Magdanz, G. Stoychev, L. Ionov, S. Sanchez, O. G. Schmidt, *Angew. Chemie - Int. Ed.* **2014**, *53*, 2673.

[26] S. Sanchez, A. A. Solovev, S. Schulze, O. G. Schmidt, *Chem. Commun.* **2011**, *47*, 698.

[27] A. A. Solovev, S. Sanchez, M. Pumera, Y. F. Mei, O. C. Schmidt, *Adv. Funct. Mater.* **2010**, *20*, 2430.

[28] S. Balasubramanian, D. Kagan, C. M. Jack Hu, S. Campuzano, M. J. Lobo-Castañon, N. Lim, D. Y. Kang, M. Zimmerman, L. Zhang, J. Wang, *Angew. Chemie - Int. Ed.* **2011**, *50*, 4161.

[29] B. Esteban-Fernández de Ávila, M. A. Lopez-Ramirez, R. Mundaca-Uribe, X. Wei, D. E. Ramírez-Herrera, E. Karshalev, B. Nguyen, R. H. Fang, L. Zhang, J. Wang, *Adv. Mater.* **2020**, *32*, 1.

[30] M. Guix, J. Orozco, M. Garcia, W. Gao, S. Sattayasamitsathit, A. Merkoçi, A. Escarpa, J. Wang, *ACS Nano* **2012**, *6*, 4445.

[31]  L. Soler, V. Magdanz, V. M. Fomin, S. Sanchez, O. G. Schmidt, *ACS Nano* **2013**, *7*, 9611.

[32]  Y. Ying, M. Pumera, *Chem. - A Eur. J.* **2019**, *25*, 106.

[33]  S. Campuzano, D. Kagan, J. Orozco, J. Wang, *Analyst* **2011**, *136*, 4621.

[34]  E. Karshalev, B. Esteban-Fernández De Ávila, J. Wang, *J. Am. Chem. Soc.* **2018**, *140*, 3810.

[35]  M. García, J. Orozco, M. Guix, W. Gao, S. Sattayasamitsathit, A. Escarpa, A. Merkoçi, J. Wang, *Nanoscale* **2013**, *5*, 1325.

[36]  R. Maria-Hormigos, B. Jurado-Sánchez, A. Escarpa, *Lab Chip* **2016**, *16*, 2397.

[37]  L. Kong, J. Guan, M. Pumera, *Curr. Opin. Electrochem.* **2018**, *10*, 174.

[38]  E. Morales-Narváez, M. Guix, M. Medina-Sánchez, C. C. Mayorga-Martinez, A. Merkoçi, *Small* **2014**, *10*, 2542.

[39]  M. Xuan, Z. Wu, J. Shao, L. Dai, T. Si, Q. He, *J. Am. Chem. Soc.* **2016**, *138*, 6492.

[40]  X. Ma, H. Feng, C. Liang, X. Liu, F. Zeng, Y. Wang, *J. Mater. Sci. Technol.* **2017**, *33*, 1067.

[41]  A. C. Hortelao, R. Carrascosa, N. Murillo-Cremaes, T. Patino, S. Sánchez, *ACS Nano* **2019**, *13*, 429.

[42]  X. Ma, S. Sanchez, *Chem. Commun.* **2015**, *51*, 5467.

[43]  D. Vilela, A. C. Hortelao, R. Balderas-Xicohténcatl, M. Hirscher, K. Hahn, X. Ma, S. Sánchez, *Nanoscale* **2017**, *9*, 13990.

[44]  X. Arqué, A. Romero-Rivera, F. Feixas, T. Patiño, S. Osuna, S. Sánchez, *Nat. Commun.* **2019**, *10*, 2826.

[45]  A. C. Hortelão, T. Patiño, A. Perez-Jiménez, À. Blanco, S. Sánchez, *Adv. Funct. Mater.* **2018**, *28*, 1705086.

[46]  A. C. Hortelão, T. Patiño, A. Perez-Jiménez, À. Blanco, S. Sánchez, *Adv. Funct. Mater.* **2017**, *1705086*, 1.

[47]  J. G. Li, G. Fornasieri, A. Bleuzen, M. Gich, A. Gloter, F. Bouquet, M. Impéror-Clerc, *Small* **2016**, *12*, 5981.

[48]  X. Ma, A. Jannasch, U.-R. Albrecht, K. Hahn, A. Miguel-López, E. Schäffer, S. Sánchez, *Nano Lett.* **2015**, *15*, 7043.

[49]  R. Golestanian, T. B. Liverpool, A. Ajdari, *New J. Phys.* **2007**, *9*, 126.

[50]  J. Parmar, D. Vilela, E. Pellicer, D. Esqué-de los Ojos, J. Sort, S. Sánchez, *Adv. Funct. Mater.* **2016**, *26*, 4152.

[51]  A. El-Ghenymy, R. M. Rodríguez, E. Brillas, N. Oturan, M. A. Oturan, *Environ. Sci. Pollut. Res.* **2014**, *21*, 8368.

[52]  X. Wang, J. Feng, Z. Zhang, W. Zeng, M. Gao, Y. Lv, T. Wei, Y. Ren, Z. Fan, *J. Colloid Interface Sci.* **2020**, *561*, 793.

# Nanorods based on mesoporous silica containing iron oxide nanoparticles as catalytic nanomotors: study of motion dynamics

Rafael Mestre[a], Núria Cadefau[a], Ana C. Hortelão[a], Jan Grzelak[b], Martí Gich[b], Anna Roig*[b], Samuel Sánchez*[a,c]

[a] R. Mestre, N. Cadefau, A. C. Hortelão, Prof. S. Sánchez
Institute for Bioengineering of Catalonia (IBEC), The Barcelona Institute of Science and Technology (BIST).
Baldiri-Reixac 10-12, 08028 Barcelona, Spain.
E-mail: ssanchez@ibecbarcelona.eu

[b] J. Grzelak, Dr. M. Gich, Prof. A. Roig.
Institut de Ciència de Materials de Barcelona (ICMAB-CSIC).
Campus UAB, 08193 Bellaterra, Catalonia, Spain.
E-mail: roig@icmab.es

[c] Prof. S. Sánchez
Institució Catalana de Recerca i Estudis Avançats (ICREA).
Passeig de Lluís Companys 23, 08010 Barcelona, Spain.
E-mail: ssanchez@ibecbarcelona.eu

## Supplementary Information

**Video S1.** Tracking of MSNR-Pt at 5% fuel concentration.

**Video S2.** Tracking of MSNR-Fe at 5% fuel concentration.

**Video S3**. Tracking of MSNR-Fe-Pt at 3% fuel concentration.

**Video S4.** Tracking of MSNR-Fe-Pt at 5% fuel concentration.

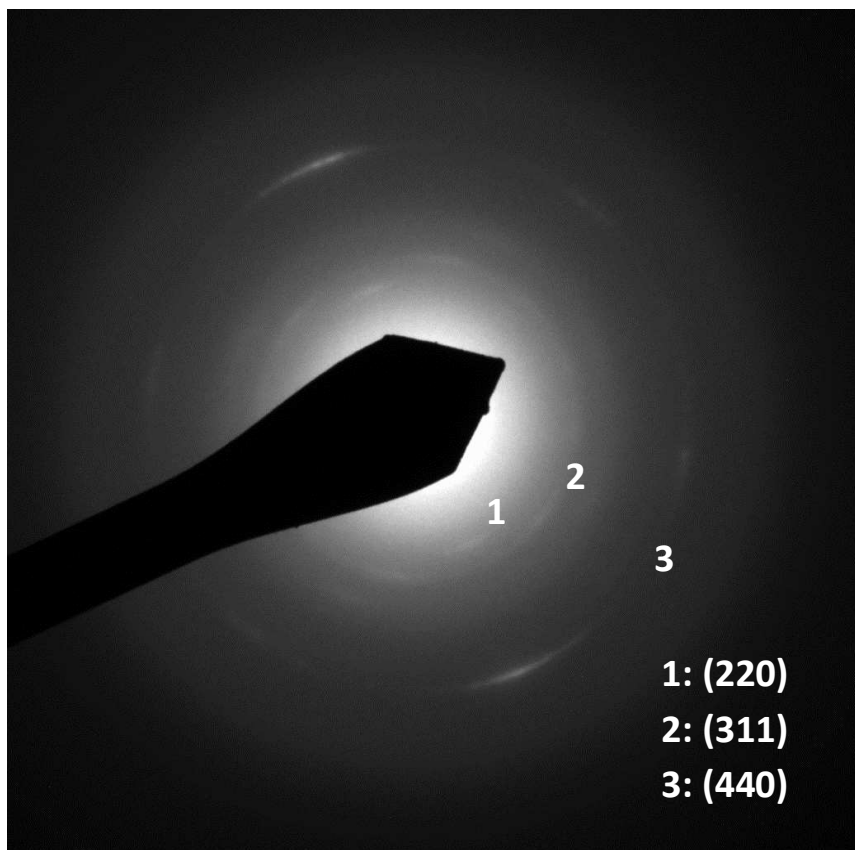

**Figure S1**. Selected area electron diffraction (SAED) pattern confirming the maghemite phase of the iron oxide nanoparticles in MSNRs-Fe.

1: (220)
2: (311)
3: (440)

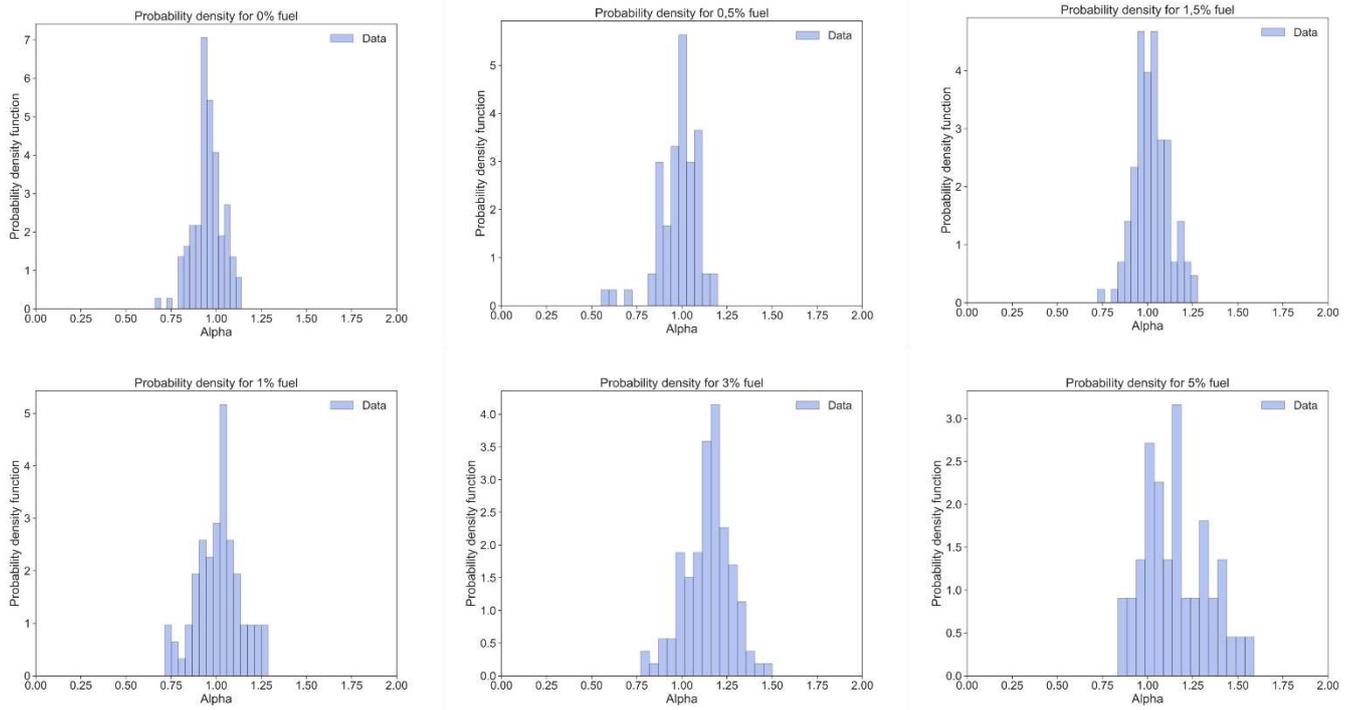

**Figure S2.** Probability density function of α for MSNR-Pt and different fuel concentrations.

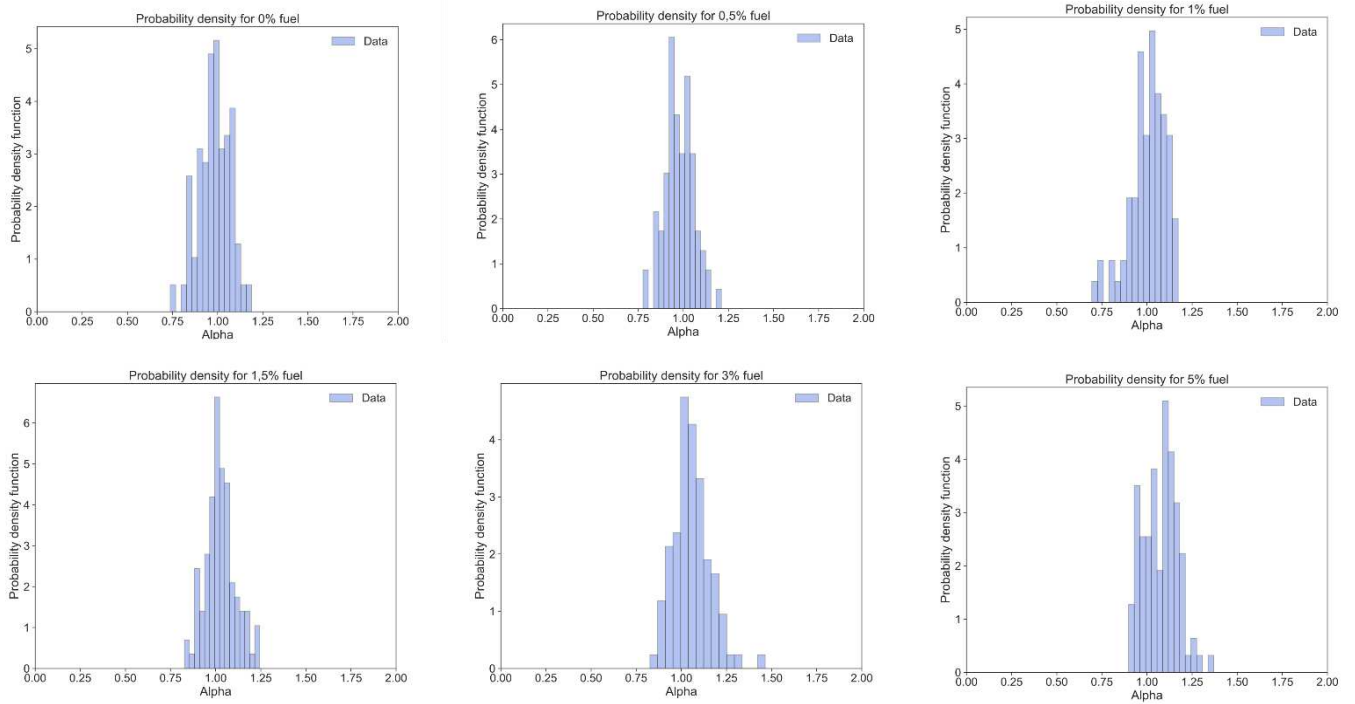

**Figure S3.** Probability density function of α for MSNR-Fe and different fuel concentrations.

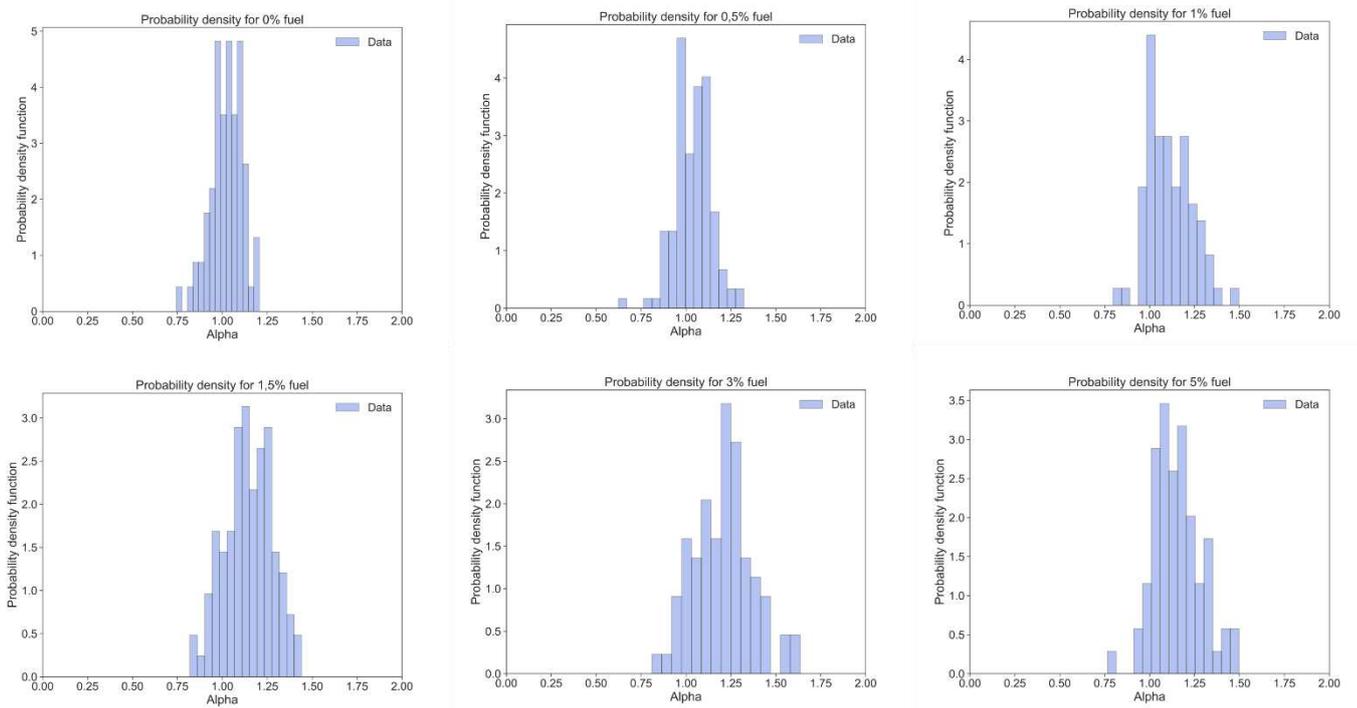

**Figure S4.** Probability density function of α for MSNR-Fe-Pt and different fuel concentrations.

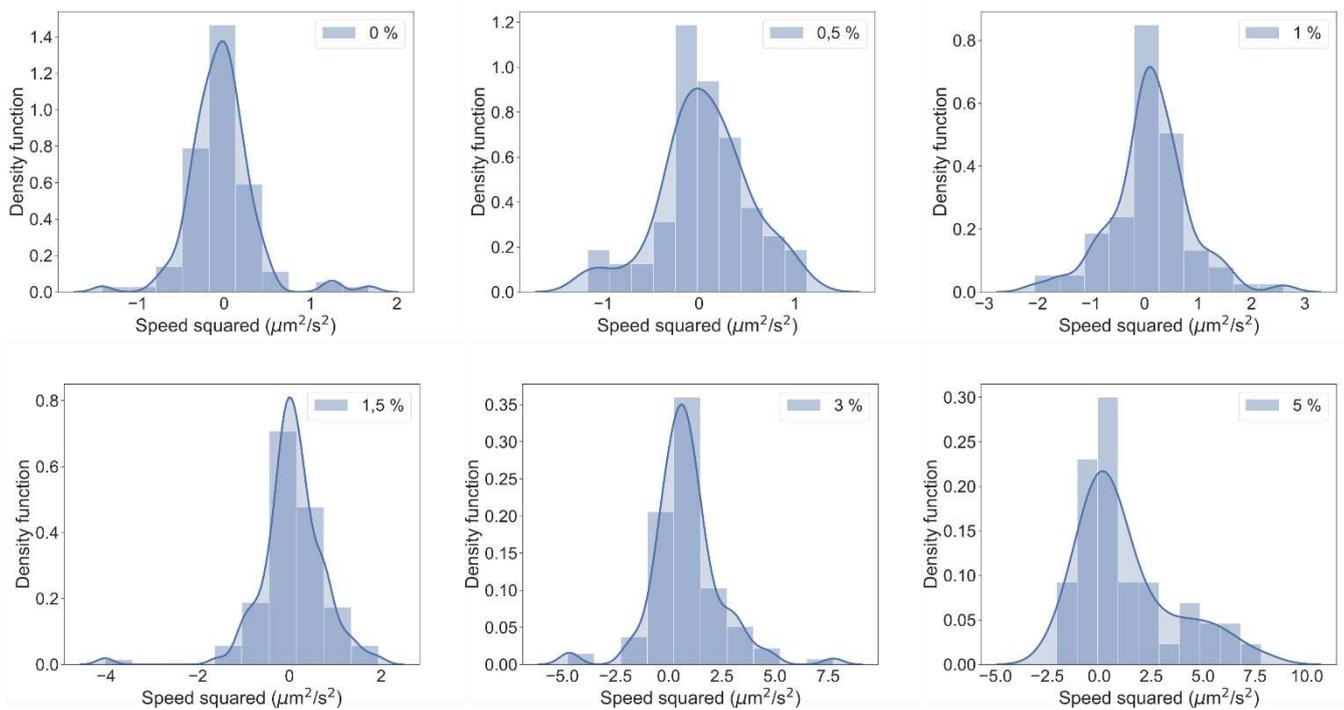

**Figure S5.** Probability density function of squared speed for MSNR-Pt and different fuel concentrations.

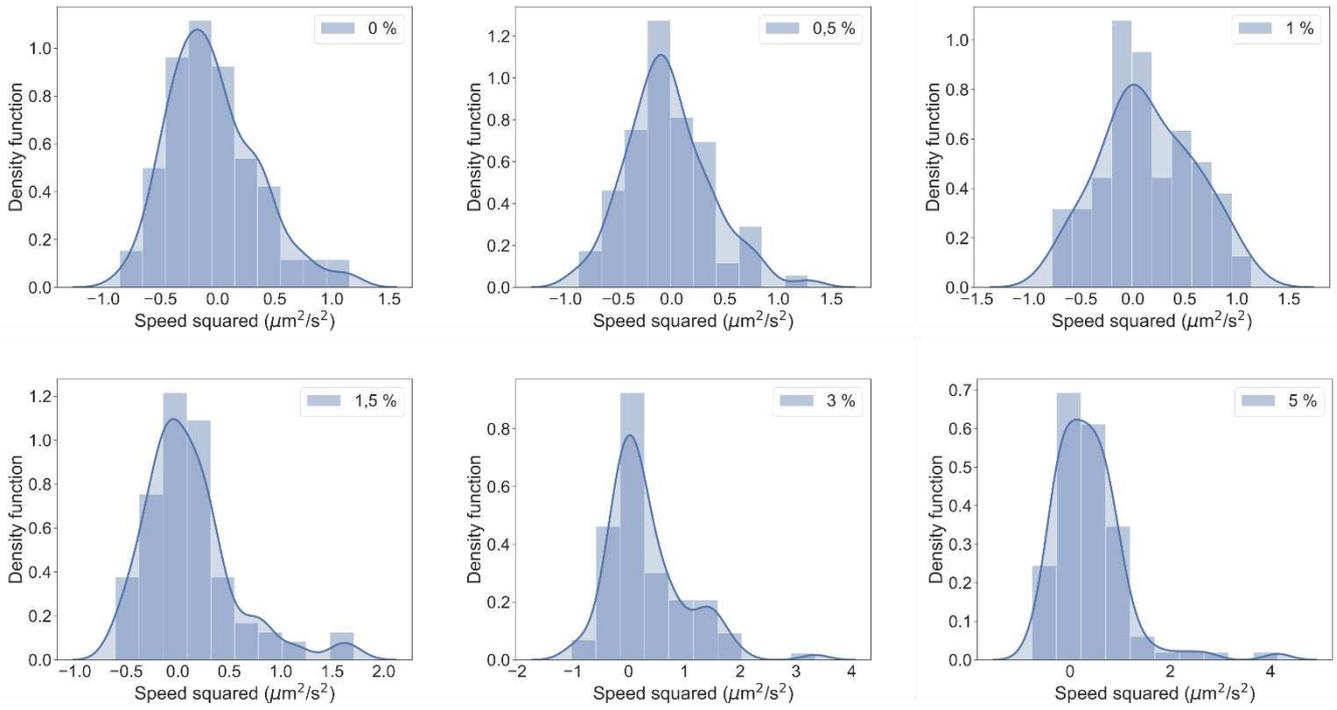

**Figure S7.** Probability density function of squared speed for MSNR-Fe and different fuel concentrations.

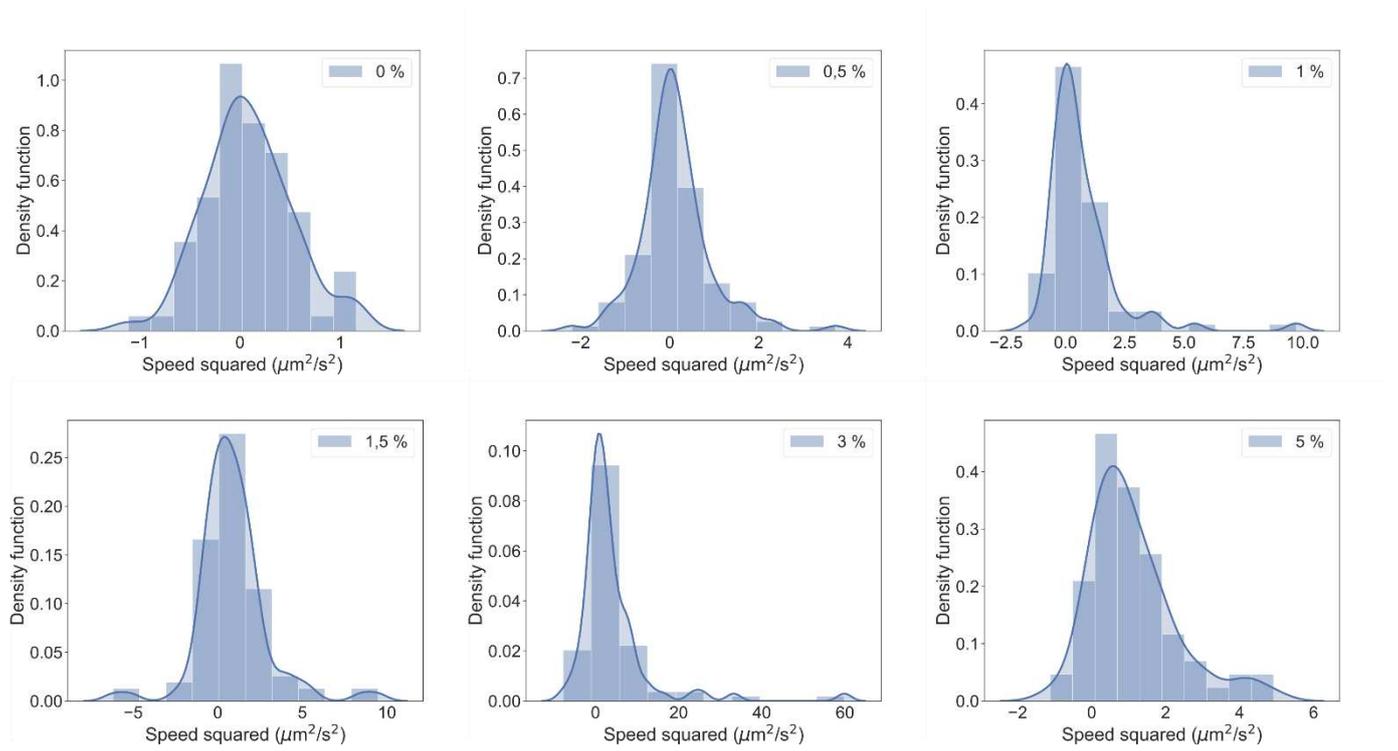

**Figure S8.** Probability density function of squared speed for MSNR-Fe-Pt and different fuel concentrations.